\documentclass[twocolumn,draft,prd]{revtex4}
\usepackage{epsf}
\usepackage{dcolumn}
\usepackage{bm}

\newcommand{\be}{\begin{equation}}
\newcommand{\ee}{\end{equation}}
\newcommand{\Deln}{\ensuremath{\Delta N_\nu\;}}
\newcommand{\epm}{\ensuremath{e^{\pm}\;}}

\def\ie{{\it i.e.},~}
\def\eg{{\it e.g.},~}
\def\etal{{\it et al.}~}
\def\4he{$^4$He}
\def\3he{$^3$He}
\def\7li{$^7$Li}
\def\Yp{Y$_{\rm P}$~}
\def\hii{H\thinspace{$\scriptstyle{\rm II}$}~}

\newcommand\la{\lower0.6ex\vbox{\hbox{\ensuremath{\buildrel{\textstyle<}\over{\sim}\ }}}}
\newcommand\ga{\lower0.6ex\vbox{\hbox{\ensuremath{\buildrel{\textstyle>}\over{\sim}\ }}}}

\newcommand{\obh}{\ensuremath{\Omega_{\rm B} h^2\;}}

\newcommand{\omb}{\ensuremath{\Omega_{\rm B}\;}}

\begin{document}

\title{BBN And CMB Constraints On Dark Energy}

\author{James P. Kneller}
\affiliation{Department of Physics, The Ohio State University,
             Columbus, OH 43210\\
             Department of Physics, North Carolina State
             University, Raleigh, NC 27695-8202~\footnote{current address}}

\author{Gary Steigman}
\affiliation{Department of Physics, The Ohio State University,
             Columbus, OH 43210~\footnote{mailing address}\\
             Department of Astronomy, The Ohio State University,
             Columbus, OH 43210}

\date{{\today}}

\begin{abstract}

Current observational data favor cosmological models which differ
from the standard model due to the presence of some form of dark
energy and, perhaps, by additional contributions to the more familiar
dark matter.  Primordial nucleosynthesis provides a window on the
very early evolution of the universe and constraints from Big Bang
Nucleosynthesis (BBN) can bound the parameters of models for dark
matter/energy at redshifts of order ten billion.  The spectrum of
temperature fluctuations imprinted on the Cosmic Microwave Background
(CMB) radiation opens a completely different window on the universe
at epochs from redshifts of order ten thousand to nearly the present.
The CMB anisotropy spectrum provides constraints on new physics which
are independent of, and complementary to those from BBN.  Here we
consider three classes of models for the dark matter/energy: extra
particles which were relativistic during the early evolution of the
universe (``$X$"); Quintessence models involving a minimally-coupled
scalar field (``$Q$"); models with a non-minimally coupled scalar
field which modify the strength of gravity during the early evolution
of the universe (``$G$").  We constrain the parameters of these models
using data from BBN and the CMB and identify the allowed regions in
their parameter spaces consistent with the more demanding joint BBN
and CMB constraints.  For ``$X$" and ``$Q$" such consistency is
relatively easy to find; it is more difficult for the ``$G$" models
with an inverse power law potential for the scalar field.

\end{abstract}


\maketitle

\section{~Introduction}

Current and ongoing spaced-based and ground-based observational
programs have provided increasingly precise data enabling us to view
the present and recent universe with hitherto unprecedented clarity and
detail.  These data have led to a current ``standard" model of
cosmology in which ``dark matter" and ``dark energy" of unknown origins
play significant roles.  At the same time, the earlier, radiation
dominated (RD) evolution of the universe remains largely hidden from
view.  These early epochs, which may harbor valuable clues to the
nature of the dark matter/energy are shrouded by the huge optical depth
of the pre-recombination plasma.  As a result, they can only be
explored indirectly, through the comparisons with observations of the
predictions of primordial nucleosynthesis (``big bang nucleosynthesis":
BBN) and of the temperature fluctuations in the spectrum of the cosmic
microwave background radiation (CMB). Studies of BBN and the CMB offer
a valuable complement to probes of the recent structure and evolution
of the universe and, in conjunction with them, may provide unique
constraints on competing models for the dark energy.  For example,
although the abundances of the light elements produced during BBN
depend largely on the universal density of baryons, they are also
sensitive to the early universe expansion rate which, in turn, is
determined by the energy density in relativistic particles and, on the
strength of gravity. Thus, BBN can not only constrain the contribution
of additional energy density beyond that predicted for the standard
model of particle physics, it can also probe the strength of the
gravitational interaction during such early epochs.  Similarly, the CMB
fluctuation spectrum depends not only on the total energy density and
the magnitude of its relativistic component, but also on the expansion
rate and the strength of gravity.  As a result, BBN and the CMB have
the potential to distinguish among -- or at least constrain --
competing models for the dark energy some of which leave the strength
of the gravity invariant while adding to the energy density, while
others may modify both.

In this paper we compare and contrast the modifications to the
standard model BBN and CMB predictions in the presence of extra,
relativistic energy (``equivalent neutrinos"), for those models
of ``Quintessence" which, during RD epochs, contribute a fixed
fraction of the relativistic energy density, and for non-minimally
coupled scalar fields (one of whose effects is to alter the
strength of the gravitational constant $G$).  After an introduction
to, and an overview of the non-standard cosmologies explored
here (\S II), the BBN predictions for these three general models
are compared and contrasted and current data is employed to provide
constraints on them in \S III. In \S IV an overview is provided
of the physical origin of the CMB fluctuations in the standard
model as a prelude to our discussion of the predictions in the
non-standard cosmologies (\S V).  The CMB constraints are presented
in \S VI and in \S VII they are combined with those from BBN
to provide joint constraints on the non-standard cosmologies
considered here.  Our results on the joint BBN-CMB constraints
are summarized in \S VIII.  Unless otherwise stated, we use
units in which $\hbar=c=8\,\pi\,G=1$.

\section{~Non-Standard Cosmologies}

Before we begin discussing the implications for BBN and the CMB of
the ``new physics" we are considering it is worth spending some
time investigating the effects of each upon cosmology in general.
As a reference we take the standard model to be geometrically
flat, containing three, light neutrinos, baryons along with cold
dark matter (CDM), and a cosmological constant ($\Lambda$).  For
numerical estimates we will often adopt the so-called concordance
values for the density parameters: $\Omega_{\rm M} = 0.3$,
$\Omega_{\Lambda} = 1 - \Omega_{\rm M} = 0.7$, and for the present
value of the Hubble parameter: H$_{0} = 72$~kms$^{-1}$Mpc$^{-1}$.

\subsection*{``$X$''}
In many extensions of the standard models of cosmology and of
particle physics there can be ``extra" energy density contained in
new particles or fields, $\rho_{X}$.  Adding energy always results
in an increase of the expansion rate, $H$, at a given redshift
since
\be H^{2} = \frac{\rho}{3} \label{eq:H2MC} \ee
where
$\rho$ is the total energy density.  The increased expansion rate
in turn implies that the age of the Universe at a given scale
factor,
\be
t(a) = \int_{0}^{a} \frac{1}{a\,H}\,da
\label{eq:age}
\ee
is younger than in the standard model.  If the
stress energy tensor for $X$ is conserved then the rate of change
of the energy density for $X$ obeys
\be \dot{\rho}_{X} + 3\,H\,(\rho_{X} +
 P_{X}) = 0
\label{eq:rhoXdot}
\ee
where the dot denotes the derivative with respect to cosmic time
and $P_{X}$ the pressure. The equation
of state, $w_{X}=P_{X}/\rho_{X}$, is therefore all that is required
to determine the evolution of $\rho_{X}$.

When $X$ behaves like radiation, that is when $w_{X}=1/3$, this
additional energy density varies with the scale factor, $a$, as
$\rho_{X} \propto a^{-4}$.  In the remainder of this paper
we shall assume this behavior for ``$X$".  Of course,
other equations of state
are possible, leading to different scalings of the X-density with
$a$ but, because this evolution of $\rho_{X}$ for $w_{X}=1/3$
is exactly the same as that for the photon and neutrino energy
densities~\footnote{Except during phases when the number of degrees
of freedom is changing such as when a massive particle species
decays or annihilates.} extra relativistic energy is a simple
extension of the standard model.

\subsection*{``$Q$''}
Another example of ``new physics" is the positing of a Quintessence
field to replace the cosmological constant as a source of the dark
energy in the standard model \cite{zws}.  Quintessence has become
an increasingly attractive alternative to $\Lambda$ because, by
making the dark energy dynamic, it helps alleviate the large
discrepancy between the matter/radiation and vacuum energy
densities most apparent during the early Universe.  Virtually all
Quintessence models are taken to be a minimally-coupled scalar
field $\phi$ with an energy density $\rho_{Q} = \dot{\phi}^{2}/2 +
V(\phi)$ and a pressure $P_{Q} = \dot{\phi}^{2}/2 - V(\phi)$ where
$V(\phi)$ is the potential energy.  If the field only interacts
with the other constituents of the cosmic fluid gravitationally
then equation (\ref{eq:rhoXdot}) also applies to $\rho_{Q}$ but
the characteristic of Quintessence is a non-trivial equation of
state that is not known a priori.  However by substituting the
expression for $\rho_{Q}$ and $P_{Q}$ into equation (\ref{eq:rhoXdot})
we obtain the Klein-Gordon equation for $\phi$
\be \ddot{\phi}
+3\,H\,\dot{\phi} + V_{\phi} = 0 \label{eq:MCevolve} \ee
where
$V_{\phi} \equiv dV/d\phi$.  Here, once $V(\phi)$ is specified, there are
no unknown functions.  For many potentials there exist solutions
to this equation to which the field converges from a wide range of
initial conditions. Solutions of this type are generically
dubbed ``tracker" solutions even though this term was originally
introduced to distinguish a specific class of models \cite{zws}.
We shall assume that the field has reached this
tracker solution long before BBN.  The evolution of the field is
controlled by the form of the potential and there are many different
models in the literature from which to choose.  This is unfortunate
because there is then no generic behavior or consequence of a dynamic
dark energy except that we may state that if the Quintessence equation
of state $w_{Q}$ is ever larger than $-1$, then the total energy
density at a given redshift must be larger than in the standard
model.  If we restrict ourselves to tracker potentials then, very
roughly, we may divide them into those where the Quintessence energy
density during the radiation dominated epoch is significant and
those where it is not. Examples of the latter type include the
popular Inverse Power Law potentials, $V \propto \phi^{-\alpha}$
\cite{fw,swz,br,ls}.  If a model is of the former type then its
equation of state must be similar to the equation of state of the
other components of the cosmic fluid $w_{f}$ and, indeed, it is
even possible during some epochs for the two to be equal.  The
Quintessence models we are considering are exactly of this type.

Any potential that fulfills the relationship $w_{Q} = w_{f}$
must reduce to the exponential potential~\footnote{The equality of
the equations of state is temporarily violated, at the $\sim 10\%$
level, during the transition from a radiation dominated to a matter
dominated fluid.} \cite{ls} $V \propto \exp(-\lambda\,\phi)$
though the Inverse Power Law potential fulfills this requirement
when $\alpha \rightarrow \infty$ as may other potentials in
appropriate limits.  This popular potential has been investigated
by Ferreira \& Joyce \cite{fj}, Liddle \& Scherrer \cite{ls}
and Copeland, Liddle \& Wands \cite{clw}, amongst others who
showed that the density parameter of the tracker solution for
the field~\footnote{The radiation/matter transition causes a
slight deviation from this formula at the $\sim 5\%$ level.}
is given by $\Omega_{Q} = 3\,(w_{f}+1)/\lambda^{2}$.  Note the
value of $\Omega_{Q}$ is larger during radiation domination than
during matter domination: $\Omega_{Q}^{(RD)} > \Omega_{Q}^{(MD)}$.
The increase in energy density results in a swifter expansion and
the age of the Universe at a fixed scale factor, again given by
equation (\ref{eq:age}), is smaller than that of the standard model
by the approximate factor $1/\sqrt{1-\Omega_{Q}}$.  However, by
itself, the exponential potential is incapable of leading to an
accelerating Universe dominated by dark energy if the Quintessence field
has reached it's tracker solution.  In order to account
for this observation the potential must depart from this simple form
as the field and the Universe evolve.  More precisely, the field must
become dominated by its potential energy in order for $w_{Q} < -1/3$.
There are many ways to achieve this result by modifying the potential
from its pure exponential form.  One modification, which we will not
discuss here, was proposed in Dodelson, Kaplinghat \& Stewart \cite{dks}.
Instead, for our analysis we have selected the Albrecht-Skordis (AS) model
\cite{as} where the potential is of the form of a product of a
polynomial (quadratic in this case) in $\phi$ and the exponential,

\be
V(\phi) = \left[ (\phi-\phi_{0})^{2} + A
 \right]\;\exp\,(-\lambda\,\phi).
\ee
The polynomial introduces a local minimum and maximum
into the potential and the parameter $A$ must satisfy $A \leq
1/\lambda^{2}$ in order for the minimum to be at a real value of
$\phi$ \cite{han}: for specificity we have chosen $A\,\lambda^{2}
= 1/100$.  With this choice for $A\,\lambda^{2}$ the two extrema
are then very close to $\phi_{0}$ and $\phi_{0}+2/\lambda$.  The
tracker solution for $\phi$ thus evolves according to the pure
exponential potential at early times when $\phi \ll \phi_{0}$.
As $\phi$ evolves and approaches the minimum at $\phi_{0}$ the
polynomial steepens the potential relative to the exponential
resulting in an increase of $\dot{\phi}$ and a simultaneous
decline in the potential energy.  This phase of $w_{Q}>w_{f}$
leads to a significant decrease in $\Omega_{Q}$.  The field passes
through $\phi_{0}$ whereupon $w_{Q}$ begins to decrease as the
field climbs out of the minimum.  When $w_{Q} = w_{f}$ again
$\Omega_{Q}$ reaches its minimum value.  If the local maximum of the potential
is sufficiently high the field is unable
to pass over the maximum at $\phi_{0}+2/\lambda$ and resume
the exponential scaling behavior, instead it becomes trapped
and begins oscillating around $\phi_{0}$ with an ever decreasing
amplitude.  After several oscillations the kinetic energy has
decreased sufficiently and $w_{Q} \sim -1$ at which point the
field begins to mimic a cosmological constant.

On a technical note, while it may appear that we have two
remaining free parameters in $V(\phi)$ after specifying
$A\,\lambda^{2}$, the solution of equation (\ref{eq:MCevolve})
depends upon $H$ which in turn is a function of $\Omega_{M0}$ and
$H_{0}$.  If we fix the geometry of the Universe to be flat then
we have specified the energy density of the field at the present
time and so introduced a constraint and eliminated another degree
of freedom.  This leaves only one remaining free parameter which
we choose to be $\lambda$.  Then, assuming the amplitude of the
oscillations around the minimum are negligible at the present
time, the imposition of the boundary conditions leads to the
result that $\lambda\,\phi_{0} \propto -2\,\ln{\lambda}$.  Also, from our
numerical calculations, we find that the minimum value of $\Omega_{Q}$ is
proportional to $\lambda^{2}$ which implies that the redshift
at which the field begins to resemble a cosmological constant,
$z_{\Lambda}$, increases with $\lambda$.  We find that roughly $z_{\Lambda}
\approx 4.7\,\lambda^{2/3}$ or equivalently $z_{\Lambda} \approx
7.5~(\,\Omega^{(RD)}_{Q}\,)^{-1/3}$.

\subsection*{``$G$''}
A third example of ``new physics" is the more radical proposal of
a non-minimal coupling between the Ricci scalar, $R$, and a scalar
field which we will further promote to the role of a Quintessence
dark energy.  This extension therefore requires the postulation
of the form of the coupling in addition to the potential $V(\phi)$.
Again there are many different models in the literature from which
to choose each involving a different form for the coupling and the
potential.  A very interesting general feature of this
class of models is that the strength of the effective gravitational
constant during the early evolution of the universe may differ from
its current value, which is fixed by terrestrial and solar system
experiments.  For this reason, we label these models by ``$G$''.
While it is possible to derive a general
formalism for the cosmology and the evolution of perturbations
\cite{hwang1,hwang2,hwang3} there is no universal behavior and
therefore it becomes necessary to restrict ourselves to a specific
example.  The model
we have adopted here is the minimal extension of the Non-Minimally
Coupled (NMC) model investigated by Chen, Scherrer \& Steigman
\cite{css} and Baccigalupi, Matarrese \& Perrotta \cite{bmp}
amongst others.  In this model the action takes the form
\be
S = \int \,d^{4}x \,\sqrt{-g}\,\left[ \frac{F(\phi)\,R}{2} -
\frac{\phi^{;\mu}\,\phi_{;\mu}}{2} -V(\phi) + {\cal L}_{f} \right]
\label{eq:action}
\ee
with $F(\phi)=1+\xi(\phi^{2}-\phi_{0}^{2})$, $\phi_{0}$ is the
value of the field today, and $\xi$ is the coupling constant.
From the action we can define the cosmological gravitational
parameter to be $1/F$ and so the evolution of the field, and
therefore $F$, will lead to an evolution of the strength of
gravity.  The potential is taken to be the previously mentioned
Inverse Power Law $V = V_{0}\,\phi^{-\alpha}$ which is known to
be a viable Quintessence model in the minimally coupled limit
\cite{ls}. There are two distinct approaches to modifications
that arise in these scenarios: redefine the energy density and
pressure of the field leaving the cosmological equations unaltered
or, adopt the minimally coupled definitions and work with the
modified cosmological equations.
Both approaches are, of course, equivalent and here we adopt
the latter.  As a result, the Friedmann equation is modified,
becoming
\be
H^{2} + \frac{H\,\dot{F}}{F} = \frac{\rho}{3\,F}
\label{eq:H2NMC}
\ee
and, after introducing the function $E$, defined by
\be
E \equiv 1 + \frac{3\,F_{\phi}^{2}}{2\,F},
\ee
the field evolves according to the equation
\begin{eqnarray}
\ddot{\phi} & + & 3\,H\,\dot{\phi} + V_{\phi}
= \frac{F_{\phi}\,R}{2} \nonumber \\ & = &
\frac{F_{\phi}}{2\,F\,E}\,\left( \rho-3\,P +
3F_{\phi}V_{\phi} -3F_{\phi\phi}\,\dot{\phi}^{2} \right).
\label{eq:NMCevolve}
\end{eqnarray}
The presence of the right hand side of equation (\ref{eq:NMCevolve})
has been dubbed the R-boost \cite{bmp}.  Examining the right hand
side of equation (\ref{eq:NMCevolve}) we discover that the last
two terms in parentheses on the right-hand side of the equation
combine to give a contribution of
\be
F_{\phi}V_{\phi} -F_{\phi\phi}\,\dot{\phi}^{2} =
-\xi\,(1+\alpha)\,\rho_{G} -\xi\,(1-\alpha)\,P_{G}
\ee
for this potential and coupling.  Again we must enforce a
self-consistent cosmology because the Hubble parameter is
still a function of $\Omega_{M0}$ and $H_{0}$ so we must
adjust the normalization constant in the potential, $V_{0}$,
to ensure the boundary conditions are matched.

The Inverse Power Law potential in the minimally coupled limit
($\xi \rightarrow 0$) is a well known and frequently studied
Quintessence model \cite{ls}.  The energy density of the field
is negligibly small during the early Universe so that $H$ is
dominated by the radiation and matter densities until close to
the present time.  The tracker solution for $\phi$ is $\phi
\propto a^{3(w_{f}+1)/(\alpha+2)}$ and for $\alpha ={\cal O}(1)$
we can immediately see that, for the tracker solution, $\phi
\ll \phi_{0}$ during much of the evolution of the Universe.
The equation of state for the field is
\be
w_{G} = \frac{ \alpha w_{f} -2 }{\alpha+2} \label{eq:wIPL}
\ee
which is always smaller than $w_{f}$ and so the Quintessence energy
density grows relative to the matter + radiation fluid~\footnote{This
confirms our previous statement concerning that
$w_{Q} \rightarrow w_{f}$ in the limit $\alpha\rightarrow \infty$.}.
This solution is also stable in the sense that perturbations from
the tracker behavior are damped \cite{ls}.  The energy density of
the field scales as $\rho_{G} \propto a^{-3\alpha(w_{f}+1)/(\alpha+2)}$.
As we approach the present era the equation of state for the field
begins to deviate from the analytic formula in equation (\ref{eq:wIPL})
and begins to approach $w_{G} = -1$.  The emergence of $\rho_{G}$,
together with the descending $w_{G}$, launches a phase of cosmological
acceleration.   As $\alpha$ increases the redshift at which the
departure from equation (\ref{eq:wIPL}) occurs also increases but
the redshift at which the Universe begins to accelerate falls because
while the Quintessence equation of state is no longer given by
equation (\ref{eq:wIPL}) its present value is still correlated with $\alpha$
with smaller $\alpha$ leading to values of $w_{G}$ that are
closer to $-1$ \cite{sk}.  For the concordance model parameters
$\Omega_{M0}=0.3$, $\Omega_{G0}=1-\Omega_{M0}=0.7$,
$H_{0}=72\;{\rm kms^{-1}Mpc^{-1}}$, the equation of state at
the present time is larger than $-1/3$ for $\alpha \gtrsim 8$
while the requirement that the Universe be accelerating requires
$\alpha \lesssim 4$ \cite{sk}.

As with our minimally coupled case ``$Q$" we shall assume that the
field has reached its tracker solution long before BBN and so our
initial condition for $\phi$ is this limit during the radiation
dominated epoch.  This may seem a minor point but the evolution
of the field in its non-tracker state will be very different from
its behavior in the tracker solution.  We posit that the tracker
solutions for non-zero $\xi$ are the same as in the minimally
coupled limit and so we must show that the R-boost is negligibly
small at early times (\ie it diverges with $a$ at a slower rate
than $V_{\phi}$ as $a \rightarrow 0$).  The coupling $F$ is almost
constant and $F_{\phi} \propto \phi$ is very small so the scaling
of the R-boost during this phase is controlled by whether $\rho_{M}
\propto a^{-3}$ or $\rho_{G}\propto a^{-4\alpha/(\alpha+2)}$ is
the more divergent: for $\alpha<6$ it is $\rho_{M}$ otherwise
it is $\rho_{G}$.  So we see that the two terms in equation
(\ref{eq:NMCevolve}) scale as $V_{\phi} \propto a^{-4\,
(\alpha+1)/(\alpha+2)}$
while $F_{\phi}\,R \propto a^{-(2+3\,\alpha)/(\alpha+2)}$ for
$\alpha<6$, $F_{\phi}\,R \propto a^{(4-4\,\alpha)/(\alpha+2)}$
for $\alpha>6$.  Thus the derivative of the potential diverges
more rapidly than the R-boost whatever the value of $\alpha$ and
the tracker solution during radiation domination is identical to
the solution in the minimally coupled limit.  It could be argued
that $\rho_{M} \rightarrow 0$ as $a\rightarrow 0$ because there
are no non-relativistic particles at such high temperatures in
which case $R \propto a^{-4\alpha/(\alpha+2)}$ independent of
$\alpha$ but this makes no difference to our conclusion.  This
tracker solution is also stable for any value of $\xi$ as shown
by Baccigalupi, Matarrese and Perrotta  \cite{bmp} who relied
on the fact that $F$ is virtually constant in order to generalize
the result of Uzan \cite{uzan}.  During matter domination the
situation is slightly different: $\phi \propto a^{3/(\alpha+2)}$
so both $V_{\phi}$ and $F_{\phi}\,R$ scale as $a^{-3(\alpha+1)/(\alpha+2)}$
and, therefore, the tracker solution is again the minimally
coupled tracker but the normalization changes to account for
the presence of the R-boost.

Since $F_{\phi}\,R$ scales more slowly than the derivative of
the potential during the radiation dominated epoch it might be
expected that eventually an R-boost phase will occur: we can
estimate the scale factor at which the two are equal to be $a \sim
a_{eq}/\xi$ where $a_{eq}$ is the scale factor at radiation-matter
equality.  An R-boost phase will only occur if $\xi \gtrsim 1$
and, once initiated, will continue into matter domination.  If
$\xi$ is small then no R-boost phase will occur during radiation
domination and neither will it commence during the matter dominated
epoch.  R-boost phases can occur when the initial value of $\phi$
differs from the tracker value.  In this case, if we again assume
$F$ is almost constant and again approximate $R$ as $R=\rho_{M}/F$,
the solution for $\phi$ during the radiation dominated epoch is
\begin{eqnarray}
\phi = \frac{\phi_{\star}}{\beta}\,\sqrt{\frac{a_{\star}}{a}}\;
J_{1}\left( 2\,\beta\,\sqrt{\frac{a}{a_{\star}}}\,\right)\;\;\;
\;\;\;& &  {\rm for}\;\xi <0, \\
\nonumber \\
\phi = \frac{\phi_{\star}}{\beta}\,\sqrt{\frac{a_{\star}}{a}}\;
I_{1}\left(2\beta\,\sqrt{\frac{a}{a_{\star}}}\,\right)\;\;\;\;
\;\; & &  {\rm for}\;\xi >0
\end{eqnarray}
where $\beta^{2}=3\,|\xi|\,\Omega_{M\star}$ and $\phi_{\star}$ is the
value of the field at $a_{\star}$, $J_{1}$ and $I_{1}$ are the
Bessel and modified Bessel functions of the first with index $1$.
In either case, when the argument of the Bessel function is
small $\phi$ evolves as
\be
\phi \approx \phi_{\star}\,\left\{ 1 \mp \frac{\beta^{2}}{2}
\left(\frac{a}{a_{\star}}\right) + \ldots\right\},
\ee
the minus (plus) sign is for $\xi <0$ ($\xi>0$).  Note that when
$\xi < 0$ the field moves backwards.  The very slow change of
$\phi$ justifies the assumption that $F$ is almost constant but,
in contrast with the tracker solution, here the behavior arises
because $\phi$ is essentially fixed rather $\phi \ll \phi_{0}$.
During this R-boost phase $\dot{\phi} \propto 1/a$ and the
potential energy is much smaller than the kinetic energy so that
$w_{G} \rightarrow 1$ but $\rho_{G} \propto 1/a^{2}$, very different
from the minimally coupled limit.  Eventually any R-boost phase will
terminate and the field will follow the tracker solution but, if
$\phi_{\star}$ is much larger than the tracker value, there may
be a very long delay before this occurs.  The approximate scale
factor at which the R-boost phase ends is given by $a =
a_{\star}\,(\phi_{\star}/\phi(a_{\star}))^{(\alpha+2)/4}$.

With the coupling $F$ almost constant during the early Universe and the
Quintessence energy density entirely negligible equation
(\ref{eq:H2NMC}) essentially reduces to the Friedmann equation of our
standard model except for an effective gravitational strength $G' =
G/F$.  Therefore the only change is to the age of the Universe at a
given scale factor, once again given by equation (\ref{eq:age}), which
is simply rescaled by the factor $\surd F$. If $F$ were not constant,
and could be adequately described by a power law function of the scale
factor, then we would attain the circumstances investigated by Carroll
and Kaplinghat \cite{ck}.

The evolution of the field becomes more complicated as the present
epoch is approached: the R-boost is no longer negligible, $F$
starts to change noticeably and the field's energy density
becomes important. Again the equation of state for the field
begins to descend towards $w_{G} = -1$ but the exact evolution
is now also a function of $\xi$.  This can be understood from the
evolution of $\rho_{G}$ which is
\be
\dot{\rho}_{G} + 3\,H\,(\rho_{G} + P_{G}) = \frac{\dot{F}\,R}{2}.
\ee
The power source $\dot{F}\,R/2$ is proportional to $\xi$ so if
$\xi>0$ then the field gains energy relative to a field in the
minimally coupled limit if $R>0$.  The power source will become
a drain if or when $R$ switches sign as the Universe begins to
accelerate.  The increase in energy density results in an increase
in the equation of state relative to the minimal case and thus
a relative increase in $\dot{\phi}$ and $\phi$.  The increased
energy density will terminate matter domination at a higher
redshift.

Finally we impose the constraints on the model parameters from
the timing experiment using the Viking probe and from limits to
the evolution of the strength of gravity \cite{pbm}. The
first constraint, the more severe of the two, is
\be
\xi\,\phi_{0} < 0.022 \label{eq:solarlimit}
\ee
while the second limits
\be
2\,\xi\,\phi_{0}\,\dot{\phi}_{0} < 10^{-11}\;{\rm yr^{-1}}
\label{eq:Gvariation}
\ee
where $\dot{\phi}_{0}$ is time derivative of the field at the
present epoch.  As shown in Chen, Scherrer and Steigman \cite{css},
$\phi_{0}$ is very weakly dependent upon $\xi$ so that it is
essentially determined by $\alpha$ and $V_{0}$.  Therefore,
with fixed values of $\alpha$ and $V_{0}$, the limit in equation
(\ref{eq:solarlimit}) is essentially a limit on $\xi$ or, if
$F_{BBN}$ is the (almost constant) value of $F$ during BBN
and if $\phi$ during this epoch is much smaller than $\phi_{0}$,
then
\be
\xi = \frac{1-F_{BBN}}{\phi_{0}^{2}}
\ee
and the limit on $\xi$ therefore becomes a limit on $F_{BBN}$
or, equivalently, on  $G'/G$.  From our numerical calculations,
again with the concordance model parameters, we find that $G'/G$
is restricted to lie between
\begin{eqnarray}
0.976 \leq G'/G \leq 1.025 & \;\;\;\;\;\;\;\;\; & \alpha=1, \nonumber \\
0.964 \leq G'/G \leq 1.040 & \;\;\;\;\;\;\;\;\; & \alpha=2, \label{eq:Glimits} \\
0.942 \leq G'/G \leq 1.067 & \;\;\;\;\;\;\;\;\; & \alpha=4. \nonumber
\end{eqnarray}

\section{~Big Bang Nucleosynthesis}

To better appreciate the similarities and differences among the three
candidates for ``new" physics under consideration here (equivalent
neutrinos:~``$X$"; Quintessence:~``$Q$"; non-minimal coupling:~``$G$")
it will be helpful to briefly review ``standard" BBN (SBBN). To this
end the discussion may begin when the universe is a few tenths of a
second old and the temperature is a few MeV. The energy density
receives its dominant contributions from the relativistic particles
present; prior to \epm annihilation these are: CBR photons, \epm pairs,
and three flavors of neutrinos, \be \rho_{\rm R} = \rho_{\gamma} +
\rho_{e} + 3\rho_{\nu} = {43 \over 8}\rho_{\gamma}. \label{rho0} \ee At
this time ($T \sim$ few MeV) the neutrinos are beginning to decouple
from the photon -- \epm plasma and the neutron to proton ratio, crucial
for the primordial abundance of \4he, is decreasing. As the temperature
drops below $\sim 2$~MeV, the two-body collisions interconverting
neutrons and protons become too slow to maintain equilibrium and the
neutron-to-proton ratio begins to deviate from (exceeds) its
equilibrium value ($(n/p)_{eq} =~ \exp(-\Delta m/T)$). Prior to \epm
annihilation, at $T \approx 0.8$~MeV when the universe is $\approx
1$~second old, the two-body reactions regulating the $n/p$ ratio become
too slow compared to the universal expansion rate and this ratio
``freezes in", although it actually continues to decrease due to the
emerging importance of ordinary beta decay ($\tau_{n} = 885.7$~sec.).
Since there are several billion CBR photons for every nucleon (baryon),
no complex nuclei exist at these early times.

BBN begins in earnest after \epm annihilation, at $T \approx
0.08$~MeV ($t \approx 3$~minutes), when the number density of
CMB photons with enough energy to photodissociate deuterium
(those in the tail of the black body distribution) is comparable
to the baryon density.  By this time the $n/p$ ratio has further
decreased due to beta decay, limiting (mainly) the amount of
helium-4 which can be synthesized.  As a result, the predictions
of primordial nucleosynthesis depend sensitively on the early
expansion rate.  In SBBN it is assumed that the neutrinos are
fully decoupled prior to \epm annihilation and don't share in the
energy transferred from the annihilating \epm pairs to the CMB
photons.  Thus, in the post-\epm annihilation universe the photons
are hotter than the neutrinos and,
\be
\rho_{\rm R} = \rho_{\gamma} + 3\rho_{\nu} = 1.6813\rho_{\gamma}.
\ee

During these RD epochs the age and the energy density are related
by ${4 \over 3}\rho_{\rm R} t^{2} = 1$ (recall that we have chosen
units in which $8\pi G = 1$), so that the age of the universe is
known (as a function of the CMB temperature) once the particle
content ($\rho_{\rm R}$) is specified {\bf and} the strength of
the gravitational interaction ($G$) is fixed.  In the standard model,
\be
{\rm Pre-\epm annihilation}:~~t~T_{\gamma}^{2} = 0.738~{\rm MeV^{2}~s},
\label{ttpre}
\ee
\be
{\rm Post-\epm annihilation}:~~t~T_{\gamma}^{2} = 1.32~{\rm MeV^{2}~s}.
\label{ttpost}
\ee
The BBN-predicted abundances of deuterium, helium-3 and lithium
are determined by the competition between various two-body
production/destruction rates and the universal expansion rate,
while the helium-4 abundance depends most directly on the neutron
abundance at the time BBN begins. As a result, the D, $^3$He, and
Li abundances are sensitive to the post-\epm annihilation expansion
rate, while that of $^4$He depends on {\bf both} the pre- and
post-\epm annihilation expansion rates; the former determines
the ``freeze-in" and the latter the importance of beta decay.
Of course, the BBN abundances do depend on the baryon density
($\eta_{10} \equiv 10^{10}n_{\rm B}/n_{\gamma} = 274\Omega_{\rm B}h^{2}$),
so that the abundances of at least two different relic nuclei
are needed to break the degeneracy between the baryon density
and a possible non-standard expansion rate resulting from new
physics or cosmology.

\subsection{~Non-Standard BBN}

Our simplest alternative to the standard cosmology
is the scenario of extra relativistic energy denoted by ``$X$".
When $X$ is decoupled in the sense that it doesn't share in
the energy released in \epm annihilation, it is convenient to
account for this extra contribution to the standard-model energy
density by normalizing it to that of an extra,``equivalent"
neutrino flavor \cite{ssg},
\be
\rho_{X} \equiv \Delta N_{\nu}\rho_{\nu} =
{7 \over 8}\Delta N_{\nu}\rho_{\gamma}.
\label{deln}
\ee
For each such ``neutrino-like" particle (\ie a two-component
fermion), if $T_{X} = T_{\nu}$, then \Deln = 1; if $X$ is a
scalar, \Deln = 4/7. However, it may well be that $X$ has
decoupled earlier in the evolution of the universe and has
failed to profit from the heating when various other
particle-antiparticle pairs annihilated (or unstable particles
decayed).  In this case, the contribution to \Deln from each
such particle will be $< 1$ ($< 4/7$).  Since we are interested
in different models of non-standard physics resulting in
modifications to the standard model energy density and
expansion rate, henceforth this case will be identified
by a superscript, $X$; \Deln = $\Delta N^{X}_{\nu}$.

In the presence of this extra component, the pre-\epm
annihilation energy density in equation (\ref{rho0}) is modified to,
\be
(\rho_{\rm R})^{X}_{pre} = {43 \over 8}\left(1 +
{7\Delta N^{X}_{\nu} \over 43}\right)\rho_{\gamma}.
\label{rhoxpre}
\ee

The extra energy density speeds up the expansion of the universe
so that the right hand side of the time-temperature relation in
equation (\ref{ttpre}) is smaller by the square root of the factor
in parentheses in equation (\ref{rhoxpre}).
\begin{eqnarray}
S^{X}_{pre} \equiv (t/t')_{pre} & = & \left(1 + {7\Delta N^{X}_{\nu} \over 43}\right)^{1/2} \nonumber \\
& = & (1 + 0.1628\Delta N^{X}_{\nu})^{1/2}.
\label{sxpre}
\end{eqnarray}
In the post-\epm annihilation universe the extra energy density
contributed by the $X$s is diluted by the heating of the photons,
so that,
\be
(\rho_{\rm R})^{X}_{post} = 1.6813\,(1 +
0.1351\Delta N^{X}_{\nu})\rho_{\gamma},
\label{rhoxpos}
\ee
and
\be
S^{X}_{post} \equiv (t/t')_{post} =
(1 + 0.1351\Delta N^{X}_{\nu})^{1/2}.
\label{sxpos}
\ee
These relations (eqs.~\ref{rhoxpre}-\ref{sxpos}) may now
be generalized to the two other cases under consideration.

For our minimally coupled Quintessence model ``$Q$" the energy
density of the field during radiation domination (relevant to
BBN), is $\Omega^{(RD)}_{Q} = 4/\lambda^{2}$. This extra energy density
may be written in terms of an equivalent $\Delta N^{Q}_{\nu}$,
\be
\Delta N^{Q}_{\nu} \equiv {43 \over 7}\left({\Omega^{(RD)}_{Q}
\over 1 - \Omega^{(RD)}_{Q}}\right).
\ee
Since, for this Quintessence model, the scalar field contributes
the {\it same fraction} of the total (radiation) energy density
pre- and post-\epm annihilation,
\be
(\rho_{\rm R})^{Q}_{pre} = (\rho_{\rm R})^{Q}_{post} =
{43 \over 8}\left(1 + {7\Delta N^{Q}_{\nu} \over 43}\right)\rho_{\gamma},
\label{rhophipre}
\ee
and
\be
S^{Q}_{pre} = S^{Q}_{post} \equiv S^{Q}
= (1 + 0.1628\Delta N^{Q}_{\nu})^{1/2},
\label{sphipos}
\ee
Thus for this class of Quintessence models the speed-up factor
($S^{Q}$) in the universal expansion rate prior to \epm annihilation
is the {\it same} as the speed-up factor after \epm annihilation.
In comparing with the equivalent neutrino case, we see that for
\Deln = $\Delta N^{X}_{\nu} = \Delta N^{Q}_{\nu}$, the post-\epm
annihilation universe expands faster for ``$Q$" than for ``$X$";
alternately, for the {\it same} post-\epm annihilation speed-up,
$\Delta N^{Q}_{\nu} \approx 0.83\Delta N^{X}_{\nu}$.

In our non-minimally coupled Quintessence model, ``$G$", the
Quintessence energy density during BBN is entirely negligible
hence the total energy density at a given redshift is unaltered,
\be
(\rho_{\rm R})^{G} = \rho_{\rm R}
\ee
where $\rho_{\rm R}$ is given by eqs.~25 and 27 for the pre- and
post-\epm annihilation universe respectively. However, since $t
\propto (G\rho_{\rm R})^{-1/2}$,
\be
S_{pre}^{G} = S_{post}^{G} = (G'/G)^{1/2} \equiv
\left(1 + {7\Delta N_{\nu}^{G} \over 43}\right)^{1/2}.
\ee
For this case we have defined an equivalent number of extra neutrinos
by
\be
\Delta N_{\nu}^{G} \equiv {43 \over 7}\left({\Delta G \over G}\right).
\ee
As with Quintessence, the non-minimally coupled scalar fields result
in faster expansion of the post-\epm annihilation universe for the
{\it same} increase in the pre-\epm annihilation expansion rate.
Thus, for the same \Deln the effects on BBN of ``$Q$" and ``$G$"
are identical, but they do have the potential to be different from
those due to the usual example of extra energy density in the form
of equivalent neutrinos (``$X$").  However there is one important
difference between ``$Q$" and ``$G$", namely that $\Delta N^{Q}_{\nu}$
is unconstrained (at the moment) while the limits expressed in
(\ref{eq:solarlimit}) and (\ref{eq:Gvariation}) translate to
a restricted range for $\Delta N^{G}_{\nu}$.  The range for
$\Delta N^{G}_{\nu}$ is a function of $\alpha$, the exponent of
the Inverse Power Law potential and, as shown in Chen, Scherrer
and Steigman \cite{css}, the limits increase with the exponent.

\begin{figure}[htbp]
\begin{center}
\epsfxsize=3.4in
\epsfbox{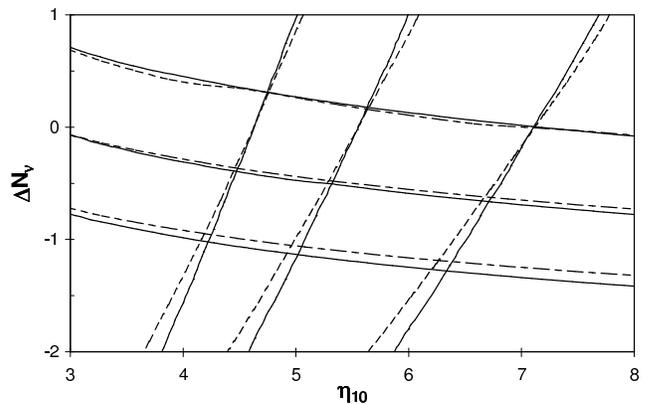}
      \caption{Isoabundance curves for D and \4he in the
      \Deln -- $\eta_{10}$ plane.  The solid curves are
      for \Deln = $\Delta N_{\nu}^{X}$, while the dashed
      curves correspond to the minimally and non-minimally
      coupled scalar field cases where \Deln = $\Delta N_{\nu}^{G}
      = \Delta N_{\nu}^{Q}$.  For \4he the curves (nearly horizontal)
      are for, from top to bottom: Y = 0.25, 0.24, 0.23.
      For D the curves (nearly vertical) are for, from
      left to right: $10^{5}$(D/H) = 4.0, 3.0, 2.0.\label{fig1}}

\end{center}
\end{figure}

On the basis of the above discussion the results of our detailed
BBN calculations may be understood.  Since the primordial
abundances of D, $^3$He, and Li freeze-in late, well after \epm
annihilation has occurred, they mainly provide a probe of
$S_{post}$.  In contrast, the $^4$He mass fraction, Y$_{\rm P}$,
retains sensitivity to {\it both} $S_{pre}$ and $S_{post}$.
Furthermore, while the abundances of D, $^3$He, and Li are most
sensitive to the baryon density, the $^4$He mass fraction provides
the best probe of the expansion rate.  This is illustrated in
Figure 1 where, in the \Deln -- $\eta_{10}$ plane are shown
isoabundance contours for D/H and Y$_{\rm P}$ (the isoabundance
curves for $^3$He/H and for Li/H, omitted for clarity, are similar
in behavior to that of D/H).  While the solid curves are for
the ``usual" extra energy density in relativistic particles,
the dashed curves show, for the same equivalent \Deln (see
eqs.~29 \& 34), the modifications to SBBN for both minimally and
non-minimally coupled scalar fields.  The trends illustrated in
Figure 1 are easy to understand in the context of our discussion
above of SBBN.  The higher the baryon density ($\eta_{10}$), the
faster primordial D is destroyed, so D/H is anticorrelated with
$\eta_{10}$.  But, the faster the universe expands (\Deln $> 0$),
the less time is available for D-destruction, so D/H is positively
correlated with $\Delta N_\nu$.  Since for the {\it same} $\Delta
N_\nu$, the post-\epm annihilation universe expands more rapidly
for ``$\phi"$ and ``$G$" than for ``$X$", the D-isoabundance
curves differ slightly as shown in Figure 1.  In contrast to D
(and to \3he and Li), the \4he mass fraction is relatively
insensitive to the baryon density, but is very sensitive to both
the pre- and post-\epm annihilation expansion rates (which control
the neutron-to-proton ratio).  The faster the universe expands,
the more neutrons are available for \4he.  Again, the effect on
Y$_{\rm P}$ of the same \Deln is slightly different for ``$X$",
than for ``$Q$" and ``$G$", as seen in Figure 1.

\subsection{~Primordial Abundances}

It is clear from Figure 1 that any BBN constraints on new physics
will be data-driven.  While D (and/or \3he and/or Li) largely
constrain the baryon density and \4he plays the same role for
$\Delta N_\nu$, there is an interplay between $\eta_{10}$ and
\Deln which is quite sensitive to the adopted abundances.  For
example, a lower primordial D/H increases the BBN-inferred value
of $\eta_{10}$, leading to a higher predicted primordial \4he mass
fraction.  If the primordial \4he mass fraction derived from the
data is ``low", then a low upper bound on \Deln will be inferred.
It is, therefore, crucial that we make every effort to avoid
biasing our conclusions by underestimating the uncertainties at
present in the primordial abundances derived from the
observational data.  To this end, first of all we concentrate on
deuterium as the baryometer of choice since its observed abundance
should have only decreased since BBN \cite{els} and the deuterium
observed in the high redshift, low metallicity QSO absorption line
systems (QSOALS) should be very nearly primordial.  The post-BBN
evolution of \3he and of Li are likely more complicated, involving
competition between production, destruction, and survival.

Even so, inferring the primordial D abundance from the QSOALS has
not been without its difficulties, with some abundance claims
withdrawn or revised.  At present there are 4 -- 5 QSOALS with
reasonably firm deuterium detections \cite{bta,btb,o'm,pb,dod}.
However, when D/H is plotted as a function of metallicity or
redshift, there is significant dispersion and the data fail to
reveal the anticipated ``deuterium plateau" \cite{gs01}.
Furthermore, subsequent observations of the D'Odorico \etal
\cite{dod} QSOALS by Levshakov \etal \cite{lev01} revealed a more
complex velocity structure and led to a revised -- and uncertain
-- deuterium abundance.  This sensitivity to the often poorly
constrained velocity structure in the absorbers is also exposed
by the analyses of published QSOALS data by Levshakov and
collaborators \cite{lev1,lev2,lev3}, which lead to
consistent but somewhat higher deuterium abundances than those
inferred from ``standard" data reduction analyses.  Given this
current state of affairs we believe that while the O'Meara \etal
\cite{o'm} estimate for the primordial abundance is likely
accurate (D/H = $3.0 \times 10^{-5}$), their error estimate ($\pm
0.4 \times 10^{-5}$) may be too small.  At the risk of erring on
the side of caution, here we adopt a range which encompasses the
uncertainties referred to above: D/H = $3.0^{+1.0}_{-0.5} \times
10^{-5}$.  Although permitting a larger than usual range in baryon
density, this choice has little direct effect on the probe of new
physics (constraints on $\Delta N_{\nu}$) which is the focus of
this study.

A similarly clouded situation exists for the primordial abundance
of \4he.  At present there are two estimates for the primordial
abundance of \4he based on large, nearly independent data sets and
analyses of low-metallicity, extragalactic \hii regions.  The
``IT" \cite{itl,it} estimate of Y$_{\rm P}({\rm IT}) = 0.244 \pm
0.002$ and the ``OS" determination \cite{os,oss,fo} of Y$_{\rm
P}({\rm OS}) = 0.234 \pm 0.003$ which is nearly $3\sigma$ lower.
Recent high quality observations of a relatively metal-rich \hii
region in the Small Magellanic Cloud (SMC) by Peimbert, Peimbert,
and Ruiz (PPR) \cite{ppr} reveal an abundance Y$_{\rm SMC} =
0.2405 \pm 0.0018$. When this abundance is extrapolated to zero
metallicity, PPR find Y$_{\rm P}({\rm PPR}) = 0.2345 \pm
0.0026$, lending some support to the lower OS value. These
comparisons among different observations suggest that unaccounted
systematic errors may dominate the statistical uncertainties.
Indeed, Gruenwald, Steigman, and Viegas \cite{gsv} argue that if
unseen neutral hydrogen in the ionized helium region of the
observed \hii regions is accounted for, the IT estimate of the
primordial abundance should be reduced to Y$_{\rm P}({\rm GSV}) =
0.238 \pm 0.003$ (see also \cite{vgs,sj}). Here, we adopt this
latter estimate for the central value but, as we did with
deuterium, the uncertainty is increased in an attempt to account
for likely systematic errors: Y$_{\rm P} = 0.238 \pm 0.005$,
leading to a $95\%$ CL range, $0.228 \le $~Y$_{\rm P} \le 0.248$;
this is in agreement with the estimate adopted by Olive, Steigman,
and Walker \cite{osw} in their review of SBBN.

\subsection{~Constraints From BBN}

\begin{figure}[htbp]
\begin{center}
\epsfxsize=3.4in
\epsfbox{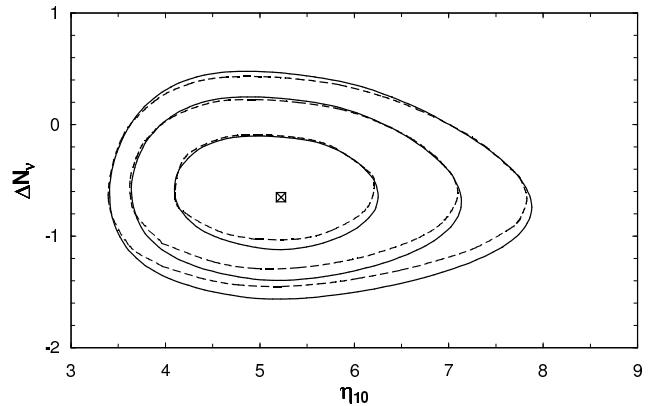}
      \caption{Likelihood contours (68\%, 95\% and 99\%
      respectively) in the \Deln -- $\eta_{10}$ plane.
      The solid curves are for \Deln = $\Delta N_{\nu}^{X}$,
      while the dashed curves correspond to the minimally
      and non-minimally coupled scalar field cases where
      \Deln = $\Delta N_{\nu}^{G} = \Delta N_{\nu}^{Q}$.
      The best fit points are indicated by the cross for
      ``$X$", and the square for ``$Q$"/``$G$". \label{fig2}}
\end{center}
\end{figure}

Using the abundance ranges adopted above, we have calculated the
\Deln -- $\eta_{10}$ likelihood contours which are shown in
Figures 2 \& 3.  A simple, semi-quantitative analysis can serve to
shed light on the detailed results shown in Figures 2 \& 3.  As
revealed in Figure 1, the baryon density is primarily fixed by
deuterium.  For D/H $= 3.0 \times 10^{-5}$, $\eta_{10} \approx
5.6$ (\omb $\approx 0.020$).  In contrast, \Deln is most sensitive
to \4he; $\Delta$Y$ \approx 0.013\Delta N_\nu$.  For $\eta_{10}
\approx 5.6$ and \Deln = 0, \Yp = 0.247.  Comparing this with the
central value (0.238) or the $2\sigma$ upper bound (0.248) inferred
from the data, it can be expected that \Deln $\approx -0.7$ and
\Deln $\la +0.1$ respectively. In fact, the detailed calculations
reveal that the ``best" value for ``$X$" is \Deln = $-0.65$, while
for ``$G/Q$", \Deln = $-0.58$; the $2\sigma$ upper bounds are
\Deln $\le +0.04$ in all three cases, corresponding to the bound
$\Omega_{Q}^{(RD)} \le 0.007$ for the Quintessence
model ($\lambda \ge 25$).  If in place of our adopted central value
and range for Y$_{\rm P}$, those from IT had been chosen, the contours
in Figures 2 \& 3 would shift upwards by \Deln $\approx 0.5$ and they
would be narrower in the vertical direction.  However, since at 95\%
CL the two estimates for \Yp agree, the upper bounds to \Deln will be
closely equal.  While we are most concerned with the BBN constraints
on $\Delta N_{\nu}$, we note in passing that the best estimate for
the baryon density is $\eta_{10} \approx 5.2$ (\obh $\approx 0.019$).

\begin{figure}[htbp]
\begin{center}
\epsfxsize=3.4in
\epsfbox{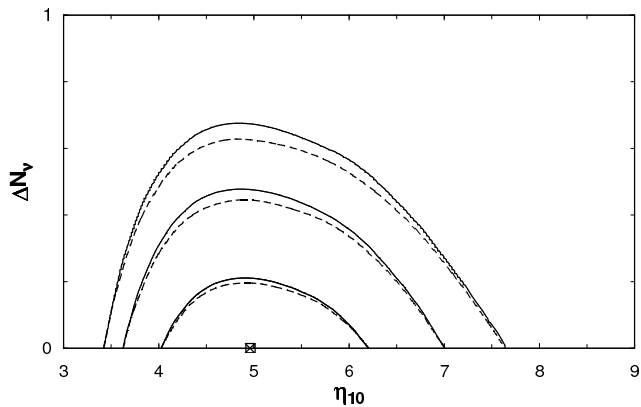}
      \caption{Likelihood contours as in Figure 2, but with the
      restriction that \Deln $\ge 0$.
      The best fit points are indicated by the cross for ``$X$",
      and the square for ``$Q$"/``$G$".\label{fig3}}
\end{center}
\end{figure}

We must point out that there is a logical inconsistency in the above
analysis.  Firstly, for two of the cases we (and others before us)
have been considering, ``$X$" and ``$Q$", the energy density and
expansion rate during BBN are only {\it increased} compared to those
for SBBN~\footnote{There is an exception provided by the type II
Randall-Sundrum model \cite{rs} which modifies the Friedmann equation
through the addition of two extra terms, one of which behaves exactly
like ``radiation" (w = 1/3) but whose sign may be positive or negative.
For a recent analysis of the constraints on such models and for further
references, see \cite{bgsw}}.  Since it is well established that there
are three, very light (hence, relativistic at BBN) neutrinos, $N_{\nu}
\ge 3$ and \Deln $\ge 0$.  Only our ``$G$" case can naturally accommodate
$S < 1$ and, hence, $\Delta N_{\nu} < 0$ (see, \eg Chen, Scherrer,
and Steigman \cite{css}).  In Figure \ref{fig3} are displayed the
corresponding likelihood contours when the restriction \Deln $\ge 0$
is imposed.  The ``best fit" now occurs for \Deln = 0 and $\eta_{10}
= 5.0$ (\obh = 0.019).  In this case the upper bounds on \Deln are
increased relative to those when \Deln is left free to be negative
as well as positive; at 95\% CL, \Deln $\le 0.43$ for ``$X$" and
\Deln $\le 0.39$ for ``$Q$".  For the Quintessence model this
corresponds to the bound $\Omega_{Q}^{(RD)} \le 0.060$ ($\lambda
\ge 8.2$).  From a similar analysis (\Deln $\ge 0$), but using
the narrower range of the IT \4he abundance as well as the O'Meara
\etal \cite{o'm} narrow deuterium range, Bean, Hansen, and Melchiorri
\cite{bean} find a 95\% upper bound of $\Omega_{Q} \le 0.045$
($\lambda \ge 9.4$).

Secondly, as mentioned before in \S II, the range of $\Delta N_{\nu}^{G}$
is constrained by the experimental limits on the present-day deviation
from GR as indicated by the allowed ranges of $G$ listed in (\ref{eq:Glimits}).
The permitted ranges of $\Delta N_{\nu}^{G}$ are, therefore, a function
of $\alpha$.  Without (yet) imposing a restriction on this parameter,
the range in \Deln is otherwise unconstrained.

As the results here demonstrate, BBN can impose significant constraints
on these non-standard models, noticeably restricting the choice
of the additional parameters which accompany the new physics.
The changes in the dynamics and evolution of the universe imposed
by these models continues after BBN, up to the present epoch.
We turn next to the effects on the formation of the CMB anisotropy
spectrum and the constraints which may be imposed on these models
by the current observational data.

\section{~Standard CMB}

In order to understand the effects of new physics upon the
formation of the anisotropies in the CMB we briefly review
the formation of temperature fluctuations in the standard
model.  More extensive reviews may be found in Ma and Bertschinger
\cite{mb}, Hu and Sugiyama \cite{hs} and elsewhere so we
will only outline the general procedure and important results.
Hereafter overdots denote derivatives with respect to conformal
time $\tau$ where $a(\tau)\,d\tau =dt$ and overbars indicate
unperturbed quantities.  We shall write the conformal Hubble
parameter as ${\cal H}$ so that $a\,{\cal H} = da/d\tau$
and therefore
\be \tau(a) = \int_{0}^{a}\,\frac{1}{a\,{\cal H}}\,da. \ee

The CMB anisotropies observed today are the redshifted temperature
fluctuations that occurred in the baryon-photon fluid prior to
recombination.
The stress energy for a perfect fluid with energy density $\rho$
and pressure $P$ is simply $T^{\mu}_{\nu}=P\,g^{\mu}_{\nu} +
(\rho+P)\,u^{\mu}\,u_{\nu}$ and the perturbations in the fluid
are introduced as $\rho = \bar{\rho} + \delta\rho=\bar{\rho}\,
(1+\delta)$, $P = \bar{P} + \delta P$ and $u^{\mu} = \bar{u}^{\mu}
+ v^{\mu}$ but also an anisotropic stress $\Sigma^{\mu\nu}=
\bar{P}\,\Pi^{\mu\nu}$ is included in order to account for shear,
viscosity and other such processes \cite{mb}.  For multiple
components the stress energy tensor is the sum of the stress
energy tensors of each component $i$, in our case vacuum
($\Lambda$), CDM (C), baryons (B), photons ($\gamma$) and
neutrinos ($\nu$), so that $\rho = \sum_{i}\rho_{i}$, etc.
For the relativistic components, such as photons and neutrinos,
the treatment as a fluid is unrealistic and, instead, the phase
space description must be used.  The temperature fluctuation,
or brightness function, $\Theta_{i} = \delta T_{i}/\bar{T}_{i}$,
is introduced for each relativistic species; the $\Theta_{i}$ are
functions of time $\tau$, position ${\bf x}$ and the direction cosines
${\bf \hat{n}}$ of the particle momenta.  There are no fluctuations
in the vacuum energy.


The manner in which scalar perturbations in the metric are introduced
defines the gauge and many different choices have appeared in the
literature.  Two of the most popular gauges, the synchronous and
the Newtonian, were compared in Ma and Bertschinger \cite{mb} and
others are discussed by many authors including Hwang \cite{hwang}.
The synchronous gauge is chosen so that the peculiar velocity of
the CDM is zero.

The perturbations in all quantities may be decomposed into their
Fourier modes (with wavenumber $k$) and, in addition, the angular
dependence in the brightness functions may expanded as an infinite
sum of Legendre polynomials with coefficients $\Theta_{\ell}$.
In terms of these moments, the thermodynamic equivalents of the
energy density, pressure, and anisotropic stress are
\be
\delta\rho_{i} = 4\,\Theta_{i0}\;\;\;\;\;\;\;\;\;v_{i}=
\Theta_{i1}\;\;\;\;\;\;\;\;\;\Pi_{i}=\frac{12}{5}\,\Theta_{i2}
\ee
for each relativistic component $i$.

All that remains is to specify the initial conditions for the
perturbations: this is accomplished by examining the behavior
on superhorizon scales deep within the radiation dominated epoch,
retaining only the growing modes, and relating the perturbations
with some `principle' such as adiabatic or isocurvature.  For
adiabatic initial conditions in the synchronous gauge we have
the well known relations \cite{mb}:
\begin{eqnarray}
\delta_{C}& = &\delta_{B}= \frac{3}{4}\,\delta_{\gamma}=
\frac{3}{4}\,\delta_{\nu} =C(k\tau)^{2}\nonumber \\
\bar{\rho}_{B}\,v_{B} & = & \frac{4}{3}\,\bar{\rho}_{\gamma}\,
v_{\gamma} = \frac{15+4\,f_{\nu}}{23+4\,f_{\nu}}\,
\bar{\rho}_{\nu}\,v_{\nu}=\frac{k\,\tau\,\delta_{C}}{9}\\
\Pi_{\nu} & = & \frac{-16\,\delta_{C}}{3\,(15+4\,f_{\nu})} \nonumber
\end{eqnarray}
where $f_{\nu}$ is the fraction of the relativistic energy
density in neutrinos and $C$ is a dimensionless constant.

The evolution of each mode is a function of its wavelength and of
the gauge.  All modes begin at superhorizon scale, $k\tau \ll 1$,
and the CDM perturbations evolve as $\delta_{C} \propto k^{2}\tau^{2}$
during both matter and radiation domination.  As time progresses
the horizon becomes bigger than the wavelengths of different modes;
modes with larger wavenumbers (smaller wavelengths) enter the horizon
before the smaller wavenumber (larger wavelength) modes.  The evolution
of $\delta_{C}$ for modes that enter the horizon during radiation
domination is stunted; they grow only as $\ln \tau$.  This stagnation
continues until matter domination whereupon the modes begin to grow
again as $\tau^{2}$.  Prior to recombination the behavior of the
baryon-photon fluid changes radically as a mode enters the sound
horizon $s_{\gamma B}$, which is given by
\be
s_{\gamma B}(\tau) = \int_{0}^{\tau}\,d\tau\,
c_{\gamma B} = \int_{0}^{a}\,\frac{da}{a\,{\cal H}}\,
\sqrt{\frac{\bar{\rho}_{\gamma}+\bar{P}_{\gamma}}{
\bar{\rho}_{\gamma}+\bar{P}_{\gamma}+\bar{\rho}_{B}}}
\label{eq:soundh}
\ee
where $c_{\gamma B}$ is the baryon-photon fluid sound speed.  Note
the presence of the baryon density in the denominator of equation
(\ref{eq:soundh}).  The perturbations present in the fluid drive
oscillations in the baryon density, the peculiar velocity, and the
photon brightness function for moments $\ell\,\in\{ 0,1\}$.  There
are no fluctuations for $\ell > 2$ because of the tight coupling
between the photons and baryons and because the isotropising effect
of Thompson scattering suppresses these moments.  The
oscillations are forced by the CDM perturbations via the metric
potentials and their amplitude increases with the ratio of the
baryon and photon densities.  Once initiated, the oscillations
continue until recombination at which point the photon-baryon/electron
interaction ceases and the photons free-stream.
The phase of each mode at recombination is recorded in the anisotropy
angular power spectrum $C_{\ell}$.  Recombination,
which occurs at $\tau_{\star},\;a_{\star}$, introduces
an important modification to the $C_{\ell}$ known as Silk
damping \cite{silk} that is due to the increase in photon mean
free path as the number of free electrons crashes.  The subsequent
increase in photon diffusion smoothes temperature fluctuations on
scales smaller than this diffusion length.  Modification of the
$\Theta_{\ell}$ does not end at recombination: the evolution of
the Universe after this time will imprint itself upon brightness
function moments through the evolution of the potentials.  This
change, known as the Integrated Sachs-Wolfe (ISW) effect, is
initiated by the fading importance of the radiation density
(the early ISW effect) and the
growing importance of the vacuum energy towards the present epoch
(the late ISW effect).  Both cause an increase in the photon
temperature variance on the scales at which these processes occur.

What emerges from the detailed calculations is the solution for
the evolution of the perturbations, for a given cosmology.  The
perturbations for each wavenumber are scaled according to an
initial power spectrum and normalized, usually to COBE.  The
photon temperature variance vector $C_{\ell}$, or ${\cal C}_{\ell}
= \ell\,(\ell+1)\,C_{\ell}/2\pi$, is then constructed.  The spectrum
of the ${\cal C}_{\ell}$ consists of a series of peaks at specific
$\ell$ and so the gross features of a model can be characterized by
the positions of the peaks, the height of the first peak relative to
the COBE normalization point and then the relative heights of all
other peaks to the first \cite{hfzt,dlsw}.  Regardless of the
curvature of the Universe, the position of the peaks is proportional to
the ratio of the comoving angular diameter distance and the size
of the sound horizon at last scattering \cite{hfzt}.  The height
of the first peak is a measure of change in the Universe since
last scattering, while the relative heights of the other peaks
and the ratios of their separations is set by the astrophysics
prior to and during recombination.  In general there is a complex
interplay between the different elements contributing to the
formation of the temperature fluctuation spectrum, but whenever
possible we shall try to couch the changes to the CMB in this
language.

\section{~Non-Standard CMB}

\subsection*{``$X$''}

The inclusion of extra relativistic energy into the CMB calculation
is relatively straightforward.  As for the photons and the three
known neutrino flavors, a brightness function is introduced that
is Fourier transformed and the angular dependence expanded in a
basis of Legendre polynomials.  If we do not allow the extra energy
density to interact with the other components (except through gravity)
then the set of equations governing the evolution of the brightness
function coefficients is exactly the same as for the neutrinos and
there is no suppression of the higher moments of the brightness
function. The initial conditions for the extra energy density
are taken to be exactly those of the neutrinos so that $N_{\nu}
\rightarrow N_{\nu} + \Delta N^{X}_{\nu}$ and $f_{\nu} \rightarrow
f_{\nu} + f_{X}$.  The effects of extra relativistic energy density
``$X$" upon the CMB anisotropy power spectrum are then equivalent
to an amplification of the neutrino sector.  The extra energy will
increase the expansion rate ${\cal H}$, with the largest changes,
relative to the standard model, occurring during the early,
radiation dominated, epoch.  The swifter expansion modifies both the
time-temperature relationship and leads to an earlier decoupling
of the photons since the scattering rate now becomes smaller than
${\cal H}$ at an earlier epoch.  Therefore both age of the Universe,
$\tau_{\star}$, and the scale factor, $a_{\star}$, at recombination
are smaller and from equation (\ref{eq:soundh}) we see that this will
lead to a smaller sound horizon, $s_{\gamma B}(\tau_{\star})$.  The
increase in ${\cal H}$ due to the extra energy density will shift
in the CMB peaks to smaller angular scales (higher $\ell$ values)
and increase the peak separation.  The extra radiation also lowers
the redshift of matter-radiation equality which will increase the
suppression of the growth of CDM perturbations on subhorizon scales
prior to matter domination and cause an increase in the temperature
variance \cite{hssw}.  The increase in the Hubble parameter also
reduces the contribution to the anisotropy spectrum from the
velocity perturbation and the Silk damping is also less effective,
further increasing the peak enhancement at large $\ell$.  Lastly,
since the extra energy density associated with ``$X$" can cluster
under the influence of gravity, increasing \Deln enhances the early
Integrated Sachs-Wolfe (ISW) effect, leading to a significant
increase in the height of the first peak relative to the others.

\subsection*{``$Q$''}

In general the perturbations in the Quintessence energy density,
pressure, and peculiar velocity, after Fourier transforming to
 k-space are
\begin{eqnarray}
\delta\rho_{Q} & = & \frac{1}{a^{2}}\,\bar{\dot{\phi}}\,
\delta\dot{\phi} + \bar{V}_{\phi}\,\delta\phi,
\label{eq:deltarho} \\
\delta P_{Q} & = & \frac{1}{a^{2}}\,\bar{\dot{\phi}}\,
\delta\dot{\phi} - \bar{V}_{\phi}\,\delta\phi,
\label{eq:deltaP} \\
v_{Q} & = & k\,\delta\phi/\bar{\dot{\phi}}
\label{eq:theta}
\end{eqnarray}
and the Klein-Gordon equation for the evolution of the perturbations
is simply
\be
\delta\ddot{\phi} + 2\,{\cal H}\,\delta\dot{\phi} +
\left(k^{2}+a^{2}\,\bar{V}_{\phi\phi}\right)\delta\phi
- \bar{\dot{\phi}}\,\dot{\delta}_{C} = 0.
\label{eq:deltaKGMC}
\ee
There is no shear term for the Quintessence field.  The density
perturbations in the field are not adiabatic \cite{sa,hu,rp}.
That is, $\delta P_{Q} - c_{Q}^{2} \delta\rho_{Q} \neq 0$ where
$c_{Q}^{2}=dP_{Q}/d\rho_{Q}$ is the adiabatic sound speed, a fact
which, according to Ratra \& Peebles \cite{rp} is ``exceedingly
fortunate" since pressure fluctuations can resist the collapse of
Quintessence density fluctuations even if the equation of state
is zero.  In principle the CMB anisotropy spectrum emerging from
a quintessence model is very different from the case of ``$X"$,
but certain effects may be missing or small depending on the exact
behavior of the field.

The changes wrought upon the CMB anisotropy spectrum from replacement
of the vacuum energy ($\Lambda$) with a Quintessence dark energy
that follows the exponential potential were investigated by Ratra
\& Peebles \cite{rp}, Ferreira \& Joyce \cite{fj} and Skordis \&
Albrecht \cite{sa}.  Deep in the radiation dominated epoch the
AS potential may be approximated by a pure exponential
and for this potential, on superhorizon scales and with adiabatic
initial conditions,  there is no change to the evolution of
perturbations, \ie $\delta_{C} \propto k^{2}\,\tau^{2}$ \cite{fj},
which is wholly expected because during this period the Quintessence
equation of state is exactly the same as that of the radiation.  The
adiabatic initial density fluctuation in $Q$ is simply \cite{fj,sa}
\be
\delta_{Q} = \frac{4}{15}\,\delta_{C} \label{eq:initMC}
\ee
which translates to a perturbation in the field \cite{sa}
\be
\delta\phi = 4\,\delta_{C}/5\lambda \;\;\;\;\;\;\;\;\;\;\;\;
\delta\dot{\phi} = 4\,\dot{\delta}_{C}/5\lambda. \label{eq:initQMC}
\ee
Though we have used equations (\ref{eq:initMC}) and (\ref{eq:initQMC})
in all our calculations the evolution of the Quintessence perturbations
is largely independent of the exact initial conditions of the field
when the equation of state is a constant \cite{dcs}.

Life becomes more interesting on subhorizon scales, $k\tau\gg 1$.  As
in Ferreira \& Joyce \cite{fj}, we take the potential derivative and
gravitational feedback terms in equation (\ref{eq:deltaKGMC}) to be
negligible relative to the $k^{2}\delta_{Q}$ piece so that, after
using $3(\bar{w}_{f}+1)-2 =2/H\tau$, we obtain
\be
\tau^{2}\,\delta\ddot{\phi} + \frac{4\,\tau}{3\,(\bar{w}_{f}+1)-2}\,
\delta\dot{\phi} + k^{2}\,\tau^{2}\,\delta\phi \approx 0.
\ee
The solutions of this equation are linear combinations of
$J_{p}(k\tau)/\tau^{p}$, $N_{p}(k\tau)/\tau^{p}$ with $p=1/2$ during
radiation domination and $p=3/2$ during matter domination \cite{fj}.
The subhorizon perturbations in the field oscillate with decaying
amplitudes and so Quintessence does not cluster.  The density contrast
of the CDM continues to grow as $\ln \tau$ during radiation domination,
but during matter domination the evolution changes to $\delta_{C}
\propto \tau^{2+\epsilon}$ where
\be
2\epsilon = 5 \sqrt{1-24\,\Omega^{(MD)}_{Q}/25} -5 \leq 0.
\ee
The lack of clustering in the Quintessence energy density, along
with its contribution to ${\cal H}$ inhibit growth in the CDM.

These results apply when the Albrecht-Skordis potential may be
approximated by a pure exponential but when the polynomial
prefactor becomes important, these results break down.  However
we can anticipate some of the effects of the field from the simple
fact that the effect of a significant Quintessence energy density
is to drive a swifter expansion.  In this regard the ``$Q$" model
we have adopted bears some similarity to ``$X$" in that the effect
of the extra energy density in the early universe leads to a smaller
sound horizon $s_{\gamma B}(\tau_{\star})$ at recombination driving
the anisotropy spectrum peaks to smaller scales and increasing their
separation.  At the same time the increase in ${\cal H}$ also leads
to a decrease in the Silk damping, thus increasing the temperature
variance at large $\ell$ relative to the first peak.  However, unlike
``$X$", Quintessence may also significantly reduce the conformal
angular diameter distance to last scattering thus partially mitigating
the shift in the location of the first CMB peak \cite{dlsw}.  For the
AS potential, this is not expected to be a large effect
because the field becomes trapped in its minimum at $z\sim z_{\Lambda}$
which, for $\Omega_{Q}^{(RD)} \lesssim 0.1$, is above $z_{\Lambda}
\gtrsim 16$.  Another major difference between ``$X$" and ``$Q$"
is the absence of clustering on subhorizon scales, hence there is
no enhanced early ISW effect for Quintessence.  But there is an ISW
effect for this potential \cite{sa} because of the dramatic decrease
in $\rho_{Q}$ as the minimum of the Quintessence potential is approached
before $\sim z_{\Lambda}$.  This results in an intermediate-to-late
ISW effect reducing all the peak amplitudes in the power spectrum
because the COBE normalization fixes the amplitude of the power
spectrum at large scales.

\subsection*{``$G$''}

Here, once again, there are two different approaches to perturbations
in the field depending upon the preference for unmodified forms of
the geometric/gravitational quantities or for unmodified stress-energy
terms.  As before, we adopt the latter approach.  From this vantage
point the perturbations in the energy density, pressure, and velocity
divergence are the same as in equations (\ref{eq:deltarho}),
(\ref{eq:deltaP}), and (\ref{eq:theta}), but the Klein-Gordon equation
for the perturbations now assumes the form
\begin{widetext}
\begin{eqnarray}
\delta\ddot{\phi} & + & 2\,{\cal H}\,\delta\dot{\phi} +
\left( k^{2}+a^{2}\,\bar{V}_{\phi\phi}\right)\delta\phi - \bar{\dot{\phi}}\,\dot{\delta}_{C}
= \frac{a^{2}}{2}\left[ \bar{F}_{\phi}\,\delta R +
\bar{F}_{\phi\phi}\,\bar{R}\,\delta\phi \right] \nonumber \\
& = & \frac{a^{2}\,\bar{R}}{2\,\bar{E}}\,
\left[ \bar{F}_{\phi\phi} -\frac{\bar{F}_{\phi}^{2}}{\bar{F}}
- \frac{3\,\bar{F}_{\phi}^{2}\,\bar{F}_{\phi\phi}}{2\,
\bar{F}} \right]\,\delta\phi  + \frac{a^{2}\,
\bar{F}_{\phi}}{2\,\bar{F}\,\bar{E}}
\left[ \delta\rho - 3\delta P + 3(\bar{F}_{\phi}\,
\bar{V}_{\phi\phi}+\bar{F}_{\phi\phi}\,\bar{V}_{\phi})\,
\delta\phi - \frac{6\,\bar{F}_{\phi\phi}\,\bar{\dot{\phi}}\,
\delta\dot{\phi}}{a^{2}} \right]
\label{eq:deltaKGNMC}
\end{eqnarray}
\end{widetext}
and we have dropped the $F_{\phi\phi\phi}$ term.  Note that for
this potential and coupling
\begin{eqnarray}
3\,(\bar{F}_{\phi}\,\bar{V}_{\phi\phi} & + & \bar{F}_{\phi\phi}\,
\bar{V}_{\phi})\,\delta\phi -  \frac{6\,\bar{F}_{\phi\phi}\,
\bar{\dot{\phi}}\,\delta\dot{\phi}}{a^{2}} \nonumber \\
& = & - 3\,\xi\,(1+\alpha)\,\delta\rho_{G} -
3\,\xi\,(1-\alpha)\,\delta P_{G}.
\end{eqnarray}
The minimally-coupled limit of this model is another frequently
studied quintessence potential \cite{bbmpv,rp,ymiko}.  This
potential is in some respects the opposite of the Albrecht-Skordis
model where the energy density during the early Universe can be
considerable and may lead to a swifter expansion prior to its
entrapment at $z_{\Lambda}$.  In contrast, for the Inverse Power
Law potential the opposite occurs: the Quintessence energy density
is inconsequential during much of the evolution of the Universe,
only becoming important as the present epoch is approached.

In the minimal-coupling limit the superhorizon perturbations during
the radiation dominated epoch may be derived in the usual way,
leading to
\be
\delta_{G} = \frac{4\,\alpha}{3\,(5\alpha+26)}\,\delta_{C}
\ee
\be
\frac{\delta\phi}{\bar{\phi}} = \frac{4}{5\alpha+26}\,\delta_{C}
\;\;\;\;\;\;\;\;\;\frac{\delta\dot{\phi}}{\bar{\dot{\phi}}} =
\frac{8+2\alpha}{5\alpha+26}\,\delta_{C}
\ee
with no change in $\delta_{C}$.  Note that when $\alpha \rightarrow
\infty$ we regain the results of the exponential potential (see
equation (\ref{eq:initMC})) and, when $\alpha =0$, the vacuum result
$\delta_{\Lambda} = 0$.  Again, on superhorizon scales there is
no change in the evolution of $\delta_{C}$ during either matter
or radiation domination and similarly, on subhorizon scales, the
Quintessence perturbations decay \cite{rp,fj} but unlike ``$Q$"
there is no suppression in the growth of matter perturbations.
The negligible Quintessence energy density also means that there
is no change in the size of the sound horizon at recombination.
There is, however, a significant decrease in the angular diameter
distance because of the contribution from the Quintessence energy
density at low redshifts that causes a shift in the peak positions
to smaller $\ell$ and decreases their separation.  This effect
increases with $\alpha$ because the equation of state and, thus,
the energy density $\rho_{G}$ are correlated with the exponent:
smaller $\alpha$ correspond to smaller $\rho_{G}$ at a fixed
redshift.  The peak heights are suppressed relative to the COBE
normalization point because the larger Quintessence energy
density terminates matter domination at an earlier epoch.  This
enhances the late ISW effect, increasing the variance on large
scales and therefore lowering the initial amplitude of the power
spectrum.  Once again, this effect increases with $\alpha$ for
exactly the same reason.

For the non-minimally coupled case, $\xi \neq 0$, the situation
is much more complex.  All the effects discussed above for the
minimally-coupled limit apply but now we must also take into
account the change in $F$.  A reduction in the strength of gravity,
$F\geq 1$, now slows the expansion, leading to a larger sound
horizon at recombination.  If $F$ remained constant there
would be a compensating change in the distance to the last
scattering surface and the peak positions would be unaffected.
However $F$ must decrease in order to attain $F=1$ at the present
time, so this distance can only be smaller than that required to
leave the peak positions unaltered.  Hence a decrease in $G$ shifts
the CMB anisotropy spectrum peaks to even smaller $\ell$.  The
changing $F$ also reduces the ratio of the amplitude of the
first peak to the others by partially cancelling the early ISW
effect.  Simultaneously, the decrease in ${\cal H}$ shifts the
peaks to larger scales due to the fact that thermal contact
between the baryon-photon fluid can be maintained down to a
lower the redshift.  The slower expansion also increases the
duration of recombination leading to more effective photon
diffusion, decreasing the temperature variance on small scales.

\section{~Constraints From The CMB}

Unlike their similarities for BBN, the three cases we are
considering here, extra relativistic energy density (``$X$"),
a minimally coupled scalar field (``$Q$"), and a non-minimally
coupled scalar field (``$G$"), all influence the formation
of the CMB anisotropy spectrum differently, albeit with some
features in common.  To explore the CMB constraints we use
modified versions of CMBFAST \cite{sz} to construct the
anisotropy spectra and vary \Deln and the baryon density
parameter $\eta_{10}$.  We compare the models to the data
from the BOOMERANG \cite{boom}, MAXIMA \cite{max}, and DASI
\cite{dasi} detectors, employing RADPACK \cite{rad} to determine
the goodness of fit and we assign the confidence level based
on $\Delta\chi^{2}$.  RADPACK also allows us to adjust the
calibration of each data set, at the cost of a $\chi^{2}$
penalty.  The current data covers the first three peaks in
the ${\cal C}_{\ell}$ with reasonable accuracy, so while
subtle effects may be missed, this approach is sufficient
to extract the gross features of each model.  Since here
we are concentrating on constraints in the \Deln -- $\eta$
plane, we have limited the priors, adopting those of the
``concordance $\Lambda$CDM" model: $\Omega_{\rm tot} = 1$,
$\Omega_{\rm M} = 0.3$, H$_{0} = 72$~kms$^{-1}$Mpc$^{-1}$,
along with no ``tilt" ($n = 1$).  As in related, earlier
work (see, \eg \cite{lp,hann,esp,kssw}) here we explore
the constraints on \Deln derived from the CMB anisotropy.

Because the CMB constraints in the non-minimally
coupled case ``$G$" differ from those for ``$X$'' and ``$Q$''
we discuss this case, $G$, separately.  We note here that
Chen and Kamionkowski~\cite{chka} have discussed how constraints
from the CMB may be used to provide a test of Brans-Dicke
cosmology, which bears some relation to our case ``G".

\begin{figure}[htbp]
\begin{center}
\epsfxsize=3.4in \epsfbox{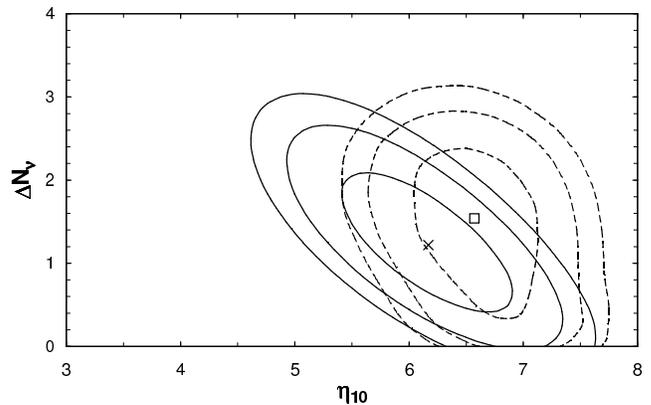}
\caption{CMB likelihood contours, 68\%, 95\%, and
      99\%, for ``$X$" (solid) and ``$Q$" (dashed).
      The best fit points are indicated by the cross
      for ``$X$", and by the square for ``$Q$".
\label{fig:CMBconstraints}}
\end{center}
\end{figure}

First, we consider ``$X$''.  In general, there is an intimate
interplay between $\eta_{10}$ and $\Delta N_{\nu}^{X}$ as shown
by the solid curves in Figure \ref{fig:CMBconstraints} where
our results reveal that an {\it increase} in \Deln can, to
some extent, be compensated by a {\it reduction} in $\eta$.
This degeneracy is the result of similar effects upon the
relative heights of the first and second peak and is only
broken by the presence of the third peak in the spectrum.
Increasing $\Delta N_{\nu}^{X}$ leads to a larger early ISW
effect boosting the variance at the first peak relative to
the rest, while increasing $\eta_{10}$ changes the relative
heights of the odd and even peaks.  At the same time
increasing $\Delta N_{\nu}^{X}$ reduces the sound horizon
by reducing the conformal age of the Universe at last
scattering, but the presence of $\bar{\rho}_{B}$ in equation
(\ref{eq:soundh}) reveals that a reduction in $\eta_{10}$
can compensate this change by increasing the baryon-photon
sound speed which overwhelms the simultaneous increase
in redshift at which recombination occurs.

The dotted curves in Figure \ref{fig:CMBconstraints} show
the contours for ``$Q$".  In this case they lack the strong
anticorrelation between \Deln and $\eta$ that ``$X$" exhibited.
Despite the differences between ``$X$" and ``$Q$", the bounds
from the CMB are quite similar in both cases, with best fit
values of $\eta_{10} \approx 6.2 - 6.5$ (\obh $\approx 0.023
- 0.024$) and non-zero \Deln $\approx 1.2 - 1.6$, while the
99\% upper bound to \Deln in both cases is \Deln $\la 3.2$
($\Omega^{(RD)}_{Q} ~\la 0.34$, $\lambda ~\ga 3.4$).

For comparison with the CMB fluctuation data, the case of non-minimally
coupled fields has several unique features which distinguish it from
the other two cases considered above.  The confidence contours shown
in Figure 5 compare the results for two choices of $\alpha$.  A
comparison of Figures 4 \& 5 reveals that the anticorrelation between
$\eta_{10}$ and $\Delta N_{\nu}^{G}$ is very strong for the non-minimally
coupled case G, with the centroid of the contours dependent on $\alpha$
as is the elongation of the contours.  We also note that the $\chi^{2}$
at the best fit point decreases with $\alpha$ \ie $\alpha=2$ is a better
fit than $\alpha=4$ and $\alpha=1$ is a better fit than $\alpha=2$.
The large shift in and the sensitivity of $\eta_{10}$ to $\alpha$
can be understood as follows.  Inverse Power Law potential models
generically {\it reduce} the heights of all the peaks because of
the late ISW effect and this suppression decreases with $\alpha$.
The peak suppression favors shifts in the experimental calibrations,
so by increasing the baryon to photon ratio there is a deflection of
some of the $\chi^{2}$ penalty by improving the fit to the first and
third peaks at the expense of making the fit to the second peak worse.
Additionally raising $\eta_{10}$ suppresses the Silk damping boosting
the temperature variance at large $\ell$ relative to that at the first
peak. Lastly, as we noted before for ``$X$", raising $\eta_{10}$ will
also reduce the sound horizon at last scattering which will compensate
for the reduction of the angular diameter distance in these models.
The shift in the centroid of the contours is due to the reduction of
the strength of the late ISW effect as $\alpha$ decreases since the
increase in $\eta_{10}$ needed to raise the peak amplitudes becomes
smaller. The anticorrelation of $\eta_{10}$ and $\Delta N_{\nu}^{G}$ is
largely the result of the change in the early ISW effect. As $\Delta
N_{\nu}^{G}$ increases the ratio of the first to second peak amplitudes
also increases which, like $\Delta N_{\nu}^{X}$, can be compensated by
reducing $\eta_{10}$.

\begin{figure}[htbp]
\begin{center}
\epsfxsize=3.4in
\epsfbox{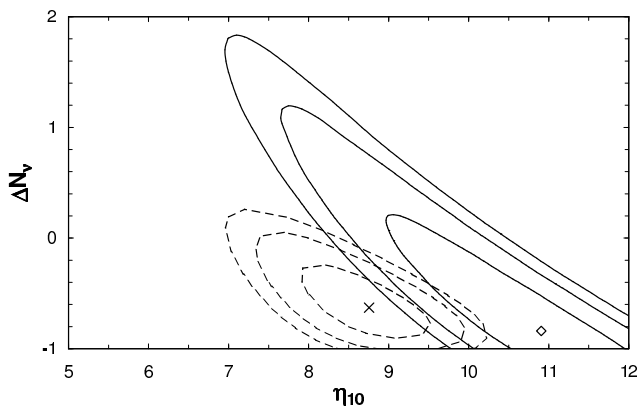}
\caption{CMB likelihood contours, 68\%, 95\%, and 99\%, for
         ``$G$" and two values of $\alpha$, the IPL exponent.
         The solid contours are for $\alpha=4$ and the dashed
         contours are for $\alpha=1$.  The best fit points are
         indicated by the diamond for $\alpha=4$  and the cross
         for $\alpha=1$.
\label{fig5}}
\end{center}
\end{figure}

Clearly, in this case the best fit values and the bounds
for $\eta_{10}$ and $\Delta N_{\nu}^{G}$ are functions of
$\alpha$.  For $\alpha=4$ we have $\eta_{10}=10.9$, $\Delta
N_{\nu}^{G}=-0.84$ with a $99\%$ upper bound of $\Delta N_{\nu}^{G}
~\la 1.8$, for $\alpha=2$ we find $\eta_{10}=10.0$,
$\Delta N_{\nu}^{G}=-0.90$ with a $99\%$ upper bound of
$\Delta N_{\nu}^{G} ~\la 0.2$ while for $\alpha=1$ we find
$\eta_{10}=8.8$, $\Delta N_{\nu}^{G}=-0.63$ with a $99\%$
upper bound of $\Delta N_{\nu}^{G} ~\la 0.3$.  However, as we
mentioned earlier, the $\chi^{2}$ at the minima for $\alpha=4$
is large (130 for 39 degrees of freedom) and we can rule out
this model as incompatible with the data an extremely high
confidence level (formally, at $\sim 10^{-11}$). The case of
$\alpha=2$ is only marginally CMB-compatible, at the 99\%
confidence level, while the $\alpha=1$ model is perfectly
acceptable.  These results are in agreement with previous
studies on limits to $\alpha$: Malquarti and Liddle \cite{ml}
find $\alpha <2$ at 95\% in the minimal limit while Bean \&
Melchiorri \cite{bm}, Hannestad \& M\"{o}rtsell \cite{hm}
and Melchiorri \etal \cite{Metal}, have found that the effective equation
of state for Quintessence must be close to $-1$ which imposes
a similar constraint on $\alpha$ \cite{sk}.

As in our discussion of BBN, we must discuss some caveats to
our CMB results.  When $\Delta N_{\nu}^{X} > 0$, the enhanced
{\it relativistic} energy density keeps the early universe
radiation-dominated to a lower redshift.  To some extent this
can be compensated by an increase in the {\it matter} density
$\Omega_{\rm M}$.  However, since in our CMB fits we have fixed
$\Omega_{\rm M}$ at 0.3, the ranges of $\Delta N_{\nu}^{X}$ shown
in Figure 4 are overly restrictive (compare, \eg with \cite{kssw}).
However, this bias will be ameliorated in the combined BBN -- CMB
likelihood distributions since, as may be seen by comparing Figures
3 and 4, the dominant constraint on $\Delta N_{\nu}^{X}$ is provided
by BBN. We must again point out that the solar system limits to $G'$
will further restrict the allowed range of $\Delta N_{\nu}^{G}$ and
indeed the best fit points shown in Figure 5 are all outside of the
allowed range corresponding to the bounds in G in eq.~(\ref{eq:Glimits}).
For non-minimally coupled scalar fields the upper and lower bounds
to $\Delta N_{\nu}^{G}$ are functions of the Inverse Power Law
exponent $\alpha$ in the adopted scalar field potential.

\section{~Combining BBN and CMB}

For a self-consistent cosmology any modifications to SBBN at redshift
$\sim 10^{10}$ must be consistent with those deviations from the
standard-model predictions for the later evolution of the universe
as probed by the CMB ($10^{4} ~\ga z ~\ga 0$).  While BBN and
the CMB favor slightly different regions in the \Deln -- $\eta$
plane, there is, indeed, overlap for models with new particles
(``$X$") and for those with minimally-coupled scalar fields
(``$Q$").  In Figure 6 are shown the joint BBN-CMB likelihood
contours for these two cases.  Although the best fits do occur for
small, nonzero values of $\Delta N_{\nu}$ (\Deln $\approx 0.1$,
$\eta_{10} \approx 6.6 \Longleftrightarrow \Omega_{\rm B}h^{2}
\approx 0.024$), the deviation from \Deln = 0 is not statistically
significant and the ``standard"  model (\Deln = 0) is entirely
consistent with the BBN and CMB data.  For both ``$X$" and ``$Q"$,
the minimum $\chi^{2}$ in our combined BBN -- CMB fit is 47 for 39
degrees of freedom; for these cases there is consistency between the
models and the universe at $z \approx 10^{10}$ and at $z \approx 0$.

\begin{figure}[htbp]
\begin{center}
\epsfxsize=3.4in
\epsfbox{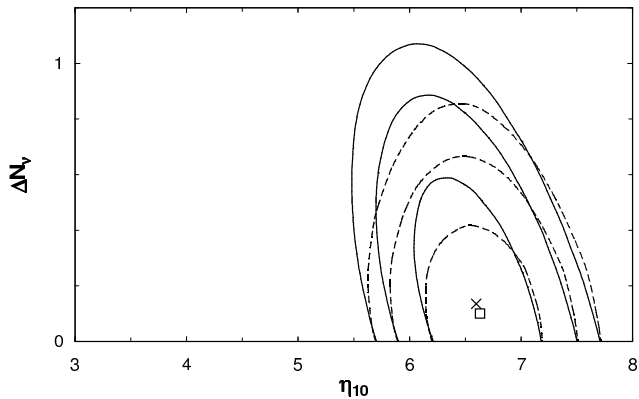}
\caption{Joint BBN and CMB likelihood contours, 68\%, 95\%, and
         99\%, for ``$X$" (solid) and ``$Q$" (dashed).  The best
         fit points are indicated by the cross for ``$X$" and by
         the square for ``$Q$".
\label{fig6}}
\end{center}
\end{figure}

In top panel of Figure 7 are shown the joint BBN-CMB likelihood
contours for the ``$G$" case, but only for $\alpha=1$: we do not
show likelihood contours for the other two $\alpha$ values since
these models ($\alpha = 2$, 4) are ruled out at better than 99\%
confidence.  For $\alpha = 1$ the $\chi^{2}$ at the minimum for
the joint BBN-CMB fits is slightly larger than for ``$X$" and
``$Q$", at 57 for 39 degrees of freedom; this non-minimally
coupled model is compatible with the data at 97\% confidence
level.  The minimum for this model occurs at $\Delta N_{\nu}^{G}
\approx -0.51$, $\eta_{10} \approx 8.2$ ($\Omega_{\rm B}h^{2}
\approx 0.030$), with a significant autocorrelation between the
two quantities.  The deviation from \Deln = 0 is now rather large:
the minimum along \Deln = 0 is located on the 90\% confidence contour.

Finally in the bottom panel of Figure 7 we impose the limits to $G'$
we have mentioned so frequently. The minimum from the top panel is
outside the allowed region $\Delta N_{\nu}^{G} = \pm 0.15$ and now
resides on the boundary at $\Delta N_{\nu}^{G}$ $\approx -0.15$,
$\eta_{10} \approx 7.6$ ($\Omega_{\rm B}h^{2} \approx 0.028$). The
$\chi^{2}$ at the minimum is such that the goodness-of-fit of the
model is only compatible with the data at the 98\% level.

\begin{figure}[htbp]
\begin{center}
\epsfxsize=3.4in \epsfbox{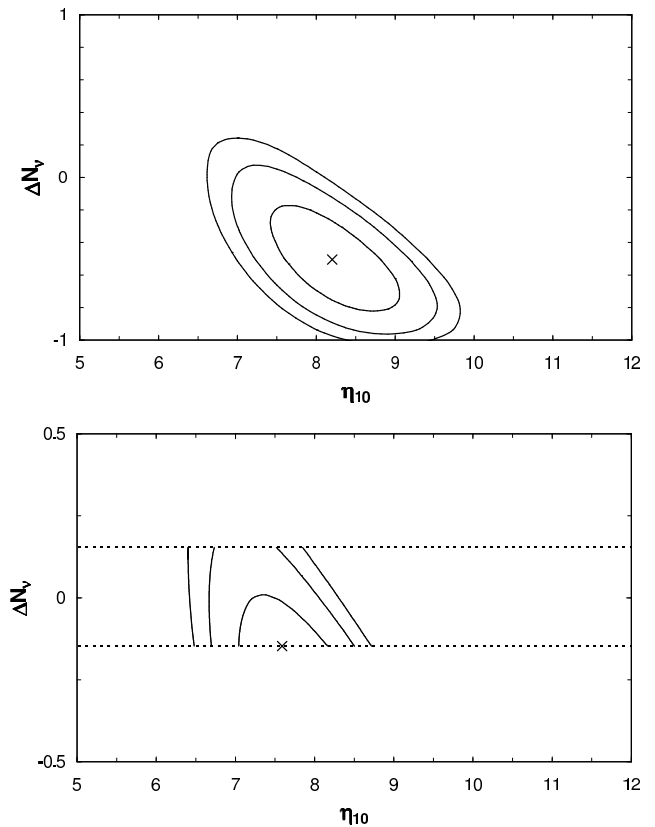}
\caption{Joint BBN and CMB likelihood contours, 68\%, 95\%,
         and 99\%, for ``$G$", $\alpha=1$. The top panel is
         with no G constraints eq.~(19), the bottom is with
         those constraints.  The best fit points are indicated
         by the crosses.
\label{fig7}}
\end{center}
\end{figure}

\section{SUMMARY}

BBN and the CMB provide complementary constraints on models
for the dark matter/energy at completely distinct epochs in
the evolution of the universe.  Consistency with both sets
of constraints can help in distinguishing among different
models for new physics beyond the standard models of cosmology
and particle physics.  As discussed here, models with extra
particles which are relativistic during the early evolution
of the universe (``$X$") and those with minimally-coupled scalar
fields (``$Q$") yield similar predictions for BBN ($\eta_{10}
\approx 5.0$ and \Deln $\approx 0$; see \S III, especially
Figs.~1 \& 3) and for the CMB ($\eta_{10} \approx 6.2 - 6.5$
and \Deln $\approx 1.2 - 1.6$; see \S VI and Fig.~4).   As a
result, they also provide similar good fits to the joint BBN-CMB
constraints (see \S VII and Fig.~6) with the baryon density being
largely determined by the CMB constraints and \Deln by BBN.
While the best fits to the data for these two options are in fact
compatible with \Deln = 0 (at $\Omega_{\rm B}h^{2} \approx 0.024$
in both cases), some dark energy is permitted: at 99\% confidence
level $\Delta N_{\nu}^{X} \la 1.07$ while $\Delta N_{\nu}^{Q} \la
0.85$ ($\Omega_{Q}^{(RD)} \la 0.12$, $\lambda \ga 5.7$).  In contrast,
the new physics associated with non-minimally coupled scalar fields
(``$G$") induces some unique behavior, leading to quite different
BBN and CMB constraints which, in general, are quite difficult to
satisfy simultaneously.  Because such models generally modify the
strength of gravity during the earlier evolution of the universe,
$\Delta_{\nu}^{G} < 0$ is not only allowed, it is favored by both
the BBN and CMB constraints (see \S III, especially Fig.~2 and
\S VI, especially Fig.~5).  However, for ``$G$" the allowed
regions in the \Deln -- $\eta$ plane are sensitive to the form
of the scalar field potential (see Fig.~5).  For inverse power
law potentials, only those models with $\alpha \la 1$ have
CMB-identified regions which have significant overlap with the
regions compatible with the BBN constraints (compare Figs.~2
\& 5).  Furthermore, since there are solar system constraints
on the possible variation of $G$, there are non-BBN and non-CMB
constraints on $G'/G$ (see eq.~(19)) which provide independent
constraints on $\Delta N_{\nu}^{G}$.  When these are combined
(for $\alpha = 1$) with the BBN and CMB constraints the best fit
value of $\Delta N_{\nu}^{G}$ is at $-0.15$ and $\Omega_{\rm B}h^{2}
\approx 0.028$ (see Fig.~7).  However, we note that there is less
than a 2\% probability that this model is, in fact, compatible
with the current BBN and CMB data.

The two models for Quintessence we have investigated in this
paper are by no means the only plausible examples of dynamic
dark energy. What distinguishes the two cases we have considered
is their importance during BBN and their role in the generation of the
primary anisotropies in the CMB.
In fact, in this regard, these
two models are unique since it is much more common for Quintessence models
to become important only at low redshifts. In such cases BBN
provides few, if any, constraints and the primary anisotropies of
the CMB are unaffected. The influence of the dark energy is via the
change in redshift of matter-dark energy equality and thus the
corresponding COBE normalization of the matter power spectrum.  For the
CMB this translates into a global stretching of the peaks in
the spectrum and a reduction in their amplitudes but for Large (and Small)
Scale Structure (LSS), Type Ia supernovae observations, and weak lensing
surveys the change in the extent of matter perturbation growth
is more important and their observation will provide valuable additional
and complimentary constraints upon the models \cite{bpb,bb,dh}.
The minimally coupled IPL potential falls into this category. The CMB already
constrains $\alpha < 2$ \cite{ml} and including LSS and Type Ia supernova data
yields $w_{G}\, \la -0.7$ at the present time which translates into $\alpha < 1.5$
for the concordance model values of $H_{0}$ and $\Omega_{M}$.

These same cosmological tests also furnish constraints upon
the two models we have examined: ``$Q$" and``$G$".
For the pure exponential potential these signatures were discussed by
Ferreira \& Joyce \cite{fj} and the Albrecht Skordis modification
(our case ``$Q$") by Skordis \& Albrect \cite{sa}. In contrast with typical
Quintessence models the AS potential modifies the matter power spectrum
mainly by inhibiting the growth of density perturbations rather than changing
the COBE normalization. The essentially stationary field below the
redshift of $z_{\Lambda} \ga 15$ (corresponding to $\lambda \ga 5.7$)
leads to little change in the magnitude-redshift relation that could
be probed by Type Ia supernovae.
For the non-minimally coupled Inverse Power Law model the effects are
supplementary to those in the minimal limit. A non-zero $\xi$ will introduce
additional effects through the change in gravitational strength $F$ that in
turn will alter the growth of density perturbations by directly influencing
their growth in addition to the change in the normalization.

\acknowledgments For valuable discussions and advice we wish to
thank R. Scherrer, M. Kaplinghat, X.Chen and A. Zentner.
We acknowledge the support of the DOE through
grant DE-FG02-91ER40690.


\end{document}